# Optical Stark decelerator for cw molecular beam with a quasi-cw semi-Gaussian laser beam

Yaling Yin, Yong Xia, and Jianping Yin*

State Key Laboratory of Precision Spectroscopy, Department of Physics, East China Normal University, Shanghai 200062, P. R. China

* Corresponding author: jpyin@phy.ecnu.edu.cn

We propose a promising scheme to decelerate a cw filtering out molecular beam using a red-detuned quasi-cw semi-Gaussian laser beam. By using Monte-Carlo simulation method, we demonstrate this optical Stark deceleration scheme, and study its decelerated dynamic process for a cw $ND_3$ molecular beam as well as the dependence of the deceleration effect on the border width of the semi-Gaussian beam. Our study shows that the proposed scheme can be used to efficiently slow a cw $ND_3$ molecular beam, and obtain a relative average kinetic-energy loss of 9.33% by using a single semi-Gaussian beam with a well depth of 7.3 mK.

It is well known that the cold dilute molecular systems play an important role in areas of cold molecular physics and cold chemistry [1,2], etc. Until now, there are two main approaches to generate slow and cold molecules. First one is to prepare cold molecules from ultracold atoms via the photoassociation [3] or Feshbach Resonance [4]. In this case, molecules are limited to only a few species of laser-cooled atoms and their short lifetime. Second, fast and hot molecules are transformed to slow and cold molecules by the methods of buffer gas cooling [5], a rotating jet [6], deceleration with a time-varying non-uniform electrostatic field [7] or magnetostatic field [8] even optical field [9], and so on. In particular, the filtering out methods based on electrostatic or magnetostatic bent guiding [10,11] have been used to generate a cw cold molecular beam with a temperature of several-ten mK from a thermal molecular ensemble with a temperature of a few K, which will be obtained by Buffer gas cooling. However, how to realize a Stark deceleration for a cw molecular beam, and how to further slow the filtering out cold molecular beam from a few 10-mK to sub-1mK? No body to date has answered these questions. So it would be interesting and worthwhile to find a useful scheme to realize further deceleration for a cw filtering out molecular beam and bring it into an ultracold regime.

In this Letter, we will use a red-detuned quasi-cw semi-Gaussian laser beam (SGB) to form a promising optical Stark decelerator, and realize an efficient deceleration for a cw filtering out (i.e., an electrostatic bend guided) molecular beam. Our proposed Stark deceleration scheme is depicted in Fig. 1, which consists of two main parts: the first part is the generation of a reduced SGB, as shown in the left of Fig. 1. A well-collimated cw Gaussian laser beam with a waist of 1.0 mm is propagated along the z-axis, and an acousto-optic modulator (AOM) is employed to modulate the cw laser beam to a quasi-cw one with a period $T$ in the time-domain. To decrease the coherence of the Gaussian laser beam and produce a neat SGB without diffracted fringes [12,13], the quasi-cw laser beam first passes through a computer-generated rotating holographic speckle (which is generated by a spatial light modulator (SLM) on a piece of liquid crystal display), and then goes through a thin and sharp semi-infinite plate that is vertically inserted to the centre of the Gaussian beam. Finally, an imaging lens is placed after the semi-infinite plate to image for the straight edge of the semi-infinite plate so as to produce a reduced SGB with a small waist of several $10-\mu m$ near the image plane of the lens. In our scheme, the transverse sizes of the SGB are reduced by *M=50* times. The intensity profile of the SGB in the *y* direction is a Gaussian one, while its intensity distribution in the *x* direction is a semi-Gaussian one. The generated reduced SGB has a waist radius of $w_{SGB}=20\mu m$ and a sharp border width $w_B=1.2\mu m$ [see Fig. 2(c)]. The second part is an optical dipole interaction between our reduced SGB and the incident cw ND$_3$ molecular beam with a temperature of 10mK, which is generated by the quadrupole electrostatic bend guiding of the buffer-gas cooled molecular source [5,10], and then an optical Stark deceleration for a cw cold molecular beam will be realized by a single SGB, and its deceleration principle will be introduced in some detail below.

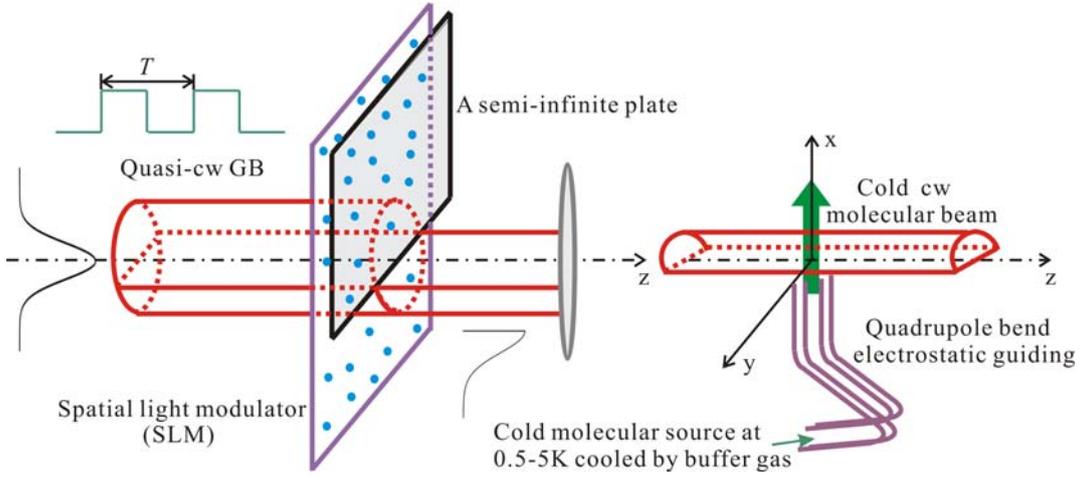

Fig. 1 A principle scheme to generate a reduced SGB and form an optical Stark decelerator for a cw molecular beam.

From ac Stark effect, the potential energy of a molecule moving in a red-detuned SGB field is given by [9]

$$U(x,y)=-\alpha I(x,y)/(2\varepsilon_0 c), \qquad (1)$$

where *I(x, y)* is the light intensity profile of the SGB, $\alpha$ is the molecular average polarizability, $\varepsilon_0$ is the dielectric constant in vacuum, and *c* the speed of light in vacuum. From $\mathbf{F}(x,y)=-\nabla U(x,y)$ and the Newton motion equation, we can obtain the following motion equations of molecules in the SGB light field:

$$\dot{v}_x=-(2\alpha I(x,y)x)/(m\varepsilon_0 c w_{SGB}^2),\quad \dot{v}_y=-(2\alpha I(x,y)y)/(m\varepsilon_0 c w_{SGB}^2). \qquad (2)$$

With these equations, we can analyze and discuss the motion of neutral molecules in the red-detuned SGB. Now let's introduce the basic deceleration principle of our SGB for a cw molecular beam. First, let's consider the interaction of a full Gaussian beam (full GB) with a cw molecular beam, as shown in Fig.2(a). For the sake of simple, we choose four typical positions (or planes) $x=x_1$, $x_2$, $x_3$ and $x_4$ to analyze the molecular motion in the light field. According to Eq. (2), when a red-detuned full GB pulse illuminates a cw incident molecular beam along the $x$ direction, the molecules at the left of $x=0$ (such as in the $x_1$ plane, etc.) will feel a dipole force parallel to their moving direction, and are accelerated during the interaction time $T/2$. Those molecules at the right of $x=0$ (such as in the $x_3$ or $x_4$ plane, etc.) will be decelerated because they experience a dipole force antiparallel to their moving direction. However, the molecules at the position $x_2$ will be first accelerated and then decelerated. As we know, the full GB is symmetric and the number of the molecules in the left of the GB is the same as that in the right. Therefore, half of molecules in the molecular beam are accelerated, while the other half is decelerated. In all, statistically the average kinetic energy of the molecular beam in the longitudinal direction keeps unchanged. So a quasi-cw full GB can't decelerate a cw molecular beam.

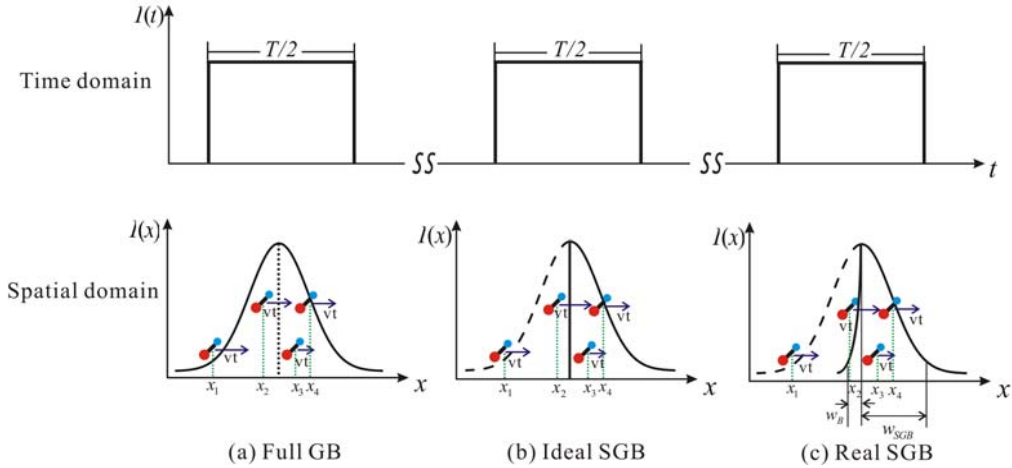

Fig. 2 Distributions of a quasi-cw Gaussian beam, an ideal and real semi-Gaussian beams in both the time and spatial domains, which will be used to describe the basic deceleration principle of our SGB scheme.

However, as shown in Fig.2(b), when a full GB pulse is replaced by an ideal SGB pulse with a border width $w_B=0$. Most of the molecules at the left of $x=0$ and far away from the $x=0$ plane (such as in the $x_1$ plane, etc.) will not be disturbed by the SGB since they are not in the SGB field at t=0 and cannot move into the SGB during $T/2$. However, only less molecules at the left of $x=0$ and near the $x=0$ plane (such as in the $x_2$ plane, etc.) will be accelerated. In this case, if a molecule moves from $x_2$ to $x'_2$ during $T/2$, its kinetic energy will be first increased by the maximum SGB well depth $-U(0)$, and then reduced by an optical Stark potential $U(x'_2)-U(0)$, here $U(x)$ are given by Eq. (1). As a result, the molecule will obtain a net kinetic-energy increase of $-U(x'_2)$. While those molecules at the right of $x=0$ (such as in the $x_3$ or $x_4$ plane, etc.) will be decelerated and will obtain a kinetic energy decrease of $U(x_3) - U(x'_3)$ or $U(x_4) - U(x'_4)$. So the cw molecular beam, to sum up, will obtain a net deceleration since the decelerated molecular number will be far greater than the accelerated one during the duration $T/2$. Since the border width $w_B (=1.2\mu m)$ of the reduced SGB is far smaller than its waist $w_{SGB}$ ($=20\mu m$), as shown in Fig.2(c), the Stark deceleration effect using a real SGB will be slightly smaller than that using an ideal SGB. Such a border width effect will be further discussed below. In addition, due to the focusing effect of the transverse dipole attractive force from the red-detuned, reduced SGB, as a molecular lens [9], the output molecular beam will be focused, and become a slow, cold and thin molecular beam.

To demonstrate our optical stark deceleration, we study the dynamic process of the slow cw $ND_3$ beam in the quasi-cw ideal SGB by using a classical Monte-Carlo simulation. In our simulation, we assume that neutral molecules in the different planes vertical to the $x$ direction have the same velocity distributions, and the initial position of the molecular beam is at $x=0$, and the molecules in the region of $0<x<v_mT/2$ are in the SGB field as the SGB comes, where $v_m$ is the mean velocity of the molecular beam. While the SGB is turned on for the time $T/2$, part of the molecules in the region $-v_mT/2<x<0$ can move into the SGB field and will be accelerated by the SGB during $T/2$. So during an AOM's modulated period $T$, the molecules in the region of $-v_mT/2<x<v_mT/2$ are disturbed. In the later periods, the above process will be repeated. In our simulation, we studied the motion process of the molecules between the $x=-v_mT/2$ and $x=v_mT/2$ planes, and we divided this simulated range into forty planes. A 1.5kW cw Ytterbium fiber laser with a wavelength of 1060-1080nm and a beam quality of $M^2<1.1$ [14] is used to generate the reduced SGB, and the corresponding maximum well depth is 7.3mK for $ND_3$, and the modulated period of AOM is set as $T=2w_{SGB}/v_m$.

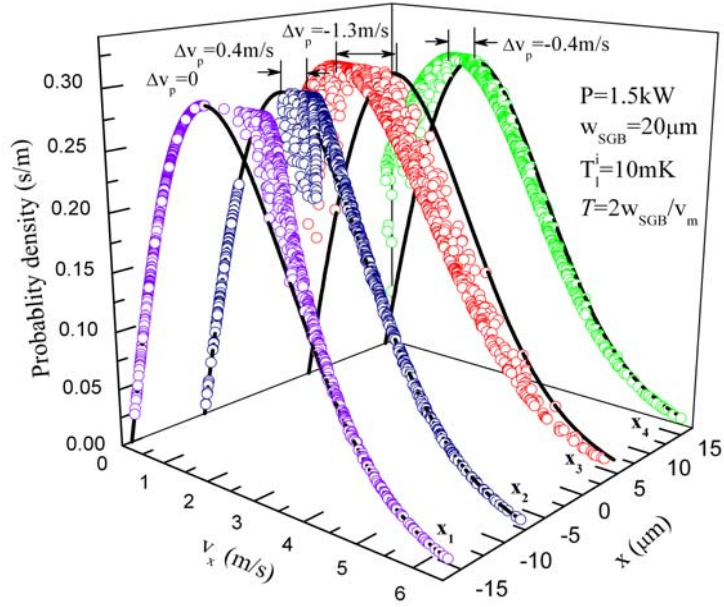

Fig. 3 The longitudinal velocity profiles of the incident $ND_3$ molecular beam in the $x$ direction and its disturbed molecules in the $x_1$, $x_2$, $x_3$ and $x_4$ planes. The solid lines represent the initial longitudinal velocity distribution of the incident $ND_3$ beam, and the circles are the Monte-Carlo simulated results that show the velocity distribution of molecules disturbed by the SGB field.

By using Monte-Carlo simulation method, we study the dynamic process of optical Stark deceleration for the $ND_3$ molecules in forty planes, and some typical simulated results are shown in Fig.3, which are four pairs of the longitudinal velocity distributions of the molecular beam in the $x=x_1$, $x=x_2$, $x=x_3$ and $x=x_4$ planes before and after the interaction with the SGB. The solid lines in Fig. 3 represent the initial longitudinal velocity distribution of the incident $ND_3$ beam, which is calculated by the Maxwell-Boltzmann's velocity distribution, while the hollow circles are the simulated velocity distributions of $ND_3$ molecules after the interaction with the SGB. As expected, a small part of the velocity distribution of the molecules in the $x=x_1$ and $x=x_2$ plane is shifted to the higher velocity direction, this shows that partial molecules in the beam were accelerated. However, the velocity shift $\Delta v_p (\equiv v_p' - v_p)$ of the most probable velocity in the $x=x_1=-0.8w_{SGB}$ plane is $\Delta v_p = 0$, while the velocity shift in the $x=x_2=-0.4w_{SGB}$ plane is $\Delta v_p = 0.4 m/s$. But the whole velocity distribution of the molecules in the $x=x_3$ and $x=x_4$ plane is shifted to the lower velocity direction, this shows that the molecular beam was decelerated. The velocity shift of the most probable velocity of molecules in the $x=x_3=0.25w_{SGB}$ plane is $\Delta v_p = -1.3 m/s$, which is far greater than the velocity shift $\Delta v_p = -0.4 m/s$ in the $x=x_4=0.8w_{SGB}$ plane. This is because when $x \geq 0$, the smaller the position $x$ of the investigated molecular plane relative to the $x=0$ plane is, the greater the obtainable optical Stark potential for molecules is. In Fig.3, there is a sharp velocity cutoff at the lower velocity extremity of the decelerated velocity profile in the $x=x_3$ and $x=x_4$ planes, this is because some much slower molecules obtain so large deceleration effect that they change their moving direction, and can not be output from the SGB. We regarded these cold molecules as a molecular loss when we plotted the simulated velocity distribution. From the above analysis, we know that the simulated results in Fig.3 are in good agreement with our introduced Stark deceleration principle [see Fig.2(b)].

We also studied the net reducing effect of the molecular kinetic energy and find that statistically the average kinetic energy of the simulated molecules is decreased by 0.933mK, corresponding to a relative energy loss of $-\Delta <E_k>/<E_k>_i = 9.33\%$. In addition, about 95% of the molecules in the incident molecular beam can be output from our reduced SGB, and the main molecular losses result from less transversely escaped molecules and those opposite-directional motional molecules. All these results show that our SGB scheme can be used to realize an efficient optical Stark deceleration for a cw molecular beam.

From the above discussions, we know that a full GB cannot be used to decelerate a cw molecular beam, while a quasi-cw ideal SGB can be used to obtain the maximum deceleration for a cw molecular beam. It is clear that a full GB corresponds to a real SGB with a border width of $w_B=w_{SGB}$, while an ideal SGB corresponds to a real SGB with a border width of $w_B=0$. Therefore, we can predict that with the increase of the border width $w_B$ from 0 to $w_{SGB}$, the deceleration effect of our SGB scheme will be reduced from the best case to zero. To verify this, we fit the intensity distribution of the real SGB in the $x$ direction with the exponential function and perform Monte-Carlo simulations for the different border width $w_B$, and the results are shown in Fig.4. In our simulation, the parameters of the SGB and molecular beam are the same as that in Fig.3. We can see from Fig.4 that when the border width of the SGB is increased from 0 to $w_B = w_{SGB} = 20\mu m$, the relative loss of the average kinetic energy of the molecular beam will be monotonically reduced from about 9.6% to 0, which is in good agreement with our theoretical predictions mentioned above. Fortunately, the border of the real SGB in our scheme is extreme sharp and its width is only $w_B = 1.2 \mu m$, its influence on the deceleration effect (as shown in Fig.4) is so small that it can be neglected. This verifies that our optical Stark decelerator for a cw molecular beam is very efficient even when the border width of the SGB is considered.

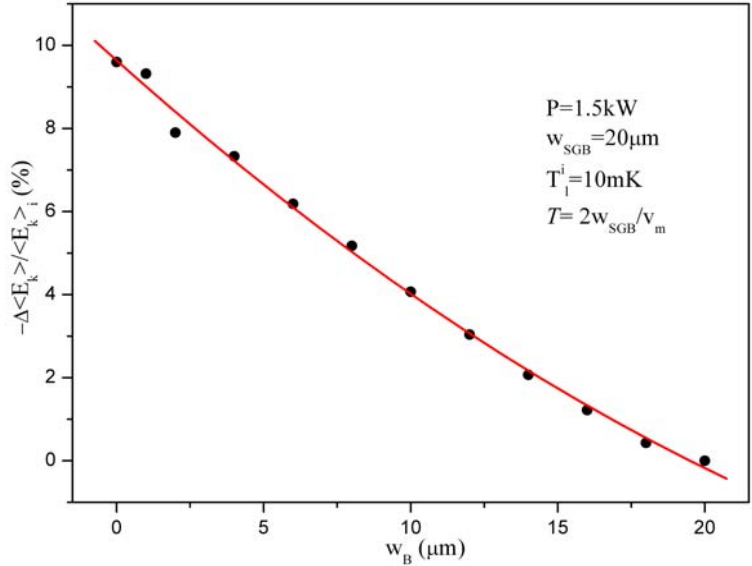

Fig. 4 Dependence of the relative average kinetic-energy loss on the border width $w_B$ of the reduced SGB. The solid circles are the Monte-Carlo simulation results, while the red solid line is the fitting curve.

In conclusions, we have proposed a promising scheme to realize further deceleration for a cw filtering out molecular beam with a red-detuned quasi-cw SGB. By using Monte-Carlo method, we have simulated the dynamic process of our optical Stark decelerator for a cw $ND_3$ molecular beam, and studied the dependence of the deceleration effect on the SGB border width $w_B$. Our study shows that the proposed scheme can be used to efficiently slow a cw $ND_3$ molecular beam, and obtain a slower and slimmer $ND_3$ molecular beam with a relative average kinetic energy loss of about 10% and a relative output molecular number of more than 90% by using a single quasi-cw SGB with a power of 1.5kW. If a cavity enhancing technique is used, the power of the SGB can be greatly lowered. In particular, when a storage ring is used to form a multistage optical Stark decelerator with a single SGB, that is, a 1D array of SGB Stark deceleration along a storage ring is formed, we can realize multiple Stark deceleration for a cw molecular beam, and bring the slowed molecules into an ultracold regime. From ac Stark effect, our scheme cannot only be used to decelerate a cw polar, non-polar or paramagnetic molecular beam, but also to decelerate a cw atomic and cluster beam, even to slow an arbitrary pulsed molecular beam. Also, our SGB cannot only be used to form various atom and molecule optical elements, such as atomic or molecular lens, atomic or molecular mirror [15], atomic or molecular grating and so on, but also to decelerate (or accelerate) some basic particles, such as electrons, plasma, etc. All these applications will be included in our future research.

This work is supported by the National Natural Science Foundation of China under Grant Nos.10374029, 10434060 and 10674047, the National Key Basic Research and Development Grogram of China under grant No.2006CB921604, the Basic Key Program of Shanghai Municipality under Grant No. 07JC14017, the Program for Changjiang Scholar and Innovative Research Team, and Shanghai Leading Academic Discipline Project under Grant No. B408.